\documentclass[%
 reprint,
preprintnumbers,
 amsmath,amssymb,
 aps,
prl,
]{revtex4-2}

\usepackage{graphicx}
\usepackage{dcolumn}
\usepackage{bm}
\usepackage[bookmarks=true, bookmarksnumbered=true, colorlinks=true, citecolor=blue, linkcolor=blue, breaklinks=true]{hyperref}

\newcommand{\RJup}{R_{\rm J}}
\newcommand{\TJup}{T_{\rm J}}

\newcommand{\GeV}{{\rm \,GeV}}
\newcommand{\cm}{{\rm \,cm}}
\newcommand{\s}{{\rm \,s}}

\newcommand{\fMB}{f_{\rm MB}}
\newcommand{\vesc}{v_{\rm esc}}
\newcommand{\nH}{n_{\rm H}}
\newcommand{\sigmav}{\langle \sigma_{\chi\chi} v \rangle}

\newcommand{\orcid}[1]{\href{https://orcid.org/#1}{\includegraphics[width=10pt]{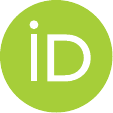}}}

\newcommand{\ror}[1]{\href{https://ror.org/#1}{\includegraphics[width=8pt]{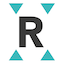}}}

\begin{document}

\preprint{KCL-PH-TH/2024-61}

\preprint{INT-PUB-24-061}

\preprint{FERMILAB-PUB-24-0820-T}

\title{Extending the Dark Matter Reach of Water Cherenkov Detectors using Jupiter}%

\author{Sandra Robles  \orcid{0000-0002-6046-8217}\,}
\email{srobles@fnal.gov}
\affiliation{Theoretical Particle Physics and Cosmology Group, Department of Physics, King’s College London, Strand, London, WC2R 2LS, UK \ror{0220mzb33}}
\affiliation{Particle Theory Department, Theory Division, Fermi National Accelerator Laboratory, Batavia, Illinois 60510, USA \ror{020hgte69}}

\author{Stephan A. Meighen-Berger  \orcid{0000-0001-6579-2000}\,}
 \email{stephan.meighenberger@unimelb.edu.au}
\affiliation{School of Physics, The University of Melbourne, Victoria 3010, Australia \ror{01ej9dk98}}
\affiliation{Center for Cosmology and AstroParticle Physics (CCAPP), Ohio State University, 
Columbus, Ohio 43210, USA \ror{00rs6vg23}}

\date{\today}

\begin{abstract}
We propose the first method for water Cherenkov detectors to constrain GeV-scale dark matter (DM) below the solar evaporation mass. While previous efforts have highlighted the Sun and Earth as DM capture targets, we demonstrate that Jupiter is a viable target. Jupiter's unique characteristics, such as its lower core temperature and significant gravitational potential, allow it to capture and retain light DM more effectively than the Sun, particularly in the mass range below 4 GeV where direct detection sensitivity diminishes. Our calculations provide the first sensitivity estimates to GeV-scale annihilating DM within Jupiter, predicting Hyper-K can reach spin dependent cross sections as low as $\sigma_{p\chi}^{\mathrm{SD}}=2\times 10^{-35}\cm^2$ for DM masses below 2 GeV. This surpasses current solar limits and direct detection results. We additionally provide estimates for Super-K ORCA, and the IceCube-Upgrade, showing that these experiments could provide complimentary bounds to direct detection experiments.
\end{abstract}

\maketitle


\textbf{\textit{Introduction.---}}
Over the past years, probing the particle-nature of GeV-scale dark matter (DM), has been believed to be the purview of direct detection experiments~\cite{Behnke:2016lsk, Amole:2019fdf, Aprile:2019xxb, Aprile:2019jmx, Aprile:2020thb, PandaX-II:2016wea, PandaX-4T:2021bab, LUX-ZEPLIN:2022xrq, CRESST:2022dtl,Lee:2024wzd, LZCollaboration:2024lux,NEWS-G:2024jms,COSINE-100:2025kbw}. These experiments rely on dark matter scattering with nucleons or electrons within the detector, detecting the energy transferred during the interaction.

In recent years, complementary methods via indirect detection of dark matter have been proposed. If DM does scatter with nucleons or electrons, it can be captured and gravitationally bound within astronomical objects. Already in the 80s, the Sun and Earth were proposed as DM capture targets~\cite{Spergel:1984re, Press:1985, Gould:1987, Gould:1987b}. Soon after, it was realized that accreted DM annihilating or decaying in the center of the Sun or the Earth could give rise to a neutrino signal potentially in the reach of neutrino detectors~\cite{Faulkner:1985rm, Silk:1985ax}. By now, this has become a well-established indirect detection technique and many neutrino experiments search for captured (and annihilating) dark matter~\cite{Super-Kamiokande:2004pou, Tanaka:2011uf, Super-Kamiokande:2015xms, ANTARES:2016xuh, IceCube:2016dgk, 
IceCube:2021xzo, Gupta:2022lws, Renzi:2023pkn}.
\begin{figure}[tb]
    \centering
    \includegraphics[width=\columnwidth]{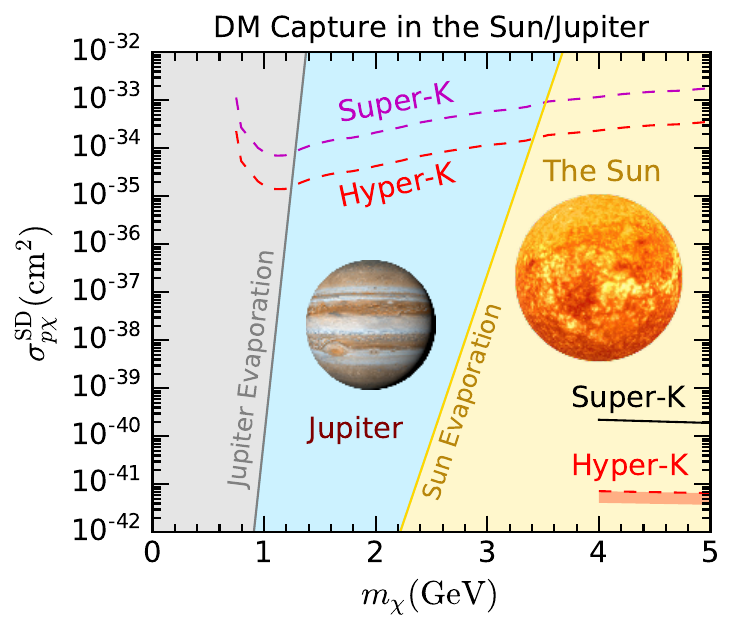}
    \caption{Sketch of the sensitivities we predicted here ($\nu\bar{\nu})$ compared to solar bounds. Super-K $\tau\tau$-bounds~\cite{Super-Kamiokande:2015xms}, Hyper-K $\nu\bar{\nu}$-estimate~\cite{Bell:2021esh}. Evaporation mass for the Sun taken from ref.~\cite{Busoni:2013kaa}. Using Jupiter, we reach energy regions \textit{inaccessible} to solar searches.}\label{fig:sketch}
\end{figure}
Indirect searches using the Sun as a target are even able to \textit{outperform} current direct detection results~\cite{Amole:2019fdf,LZCollaboration:2024lux}. In particular, this is the case for spin-dependent (SD) DM-nucleon cross sections, when the captured DM directly annihilates to neutrinos~\cite{Bell:2021esh, IceCube:2021xzo}. Such scenarios can be realized in e.g. lepton portal DM models~\cite{Okawa:2020jea, Iguro:2022tmr}. 

These indirect DM searches have intrinsic limitations. They require DM to be captured within a given star or planet, settle down in the center of the object via further scatterings and later annihilate. 
However, \textit{light} DM,  can regain energy via collisions within the astrophysical object and escape before annihilating. At which particular mass this starts to happen, depends on the celestial object. Specifically for the Sun, it has been estimated to be $\sim 4$ GeV~\cite{Griest:1986yu, Gould:1987, Hooper:2008cf, Busoni:2013kaa}. This is a similar mass to where direct detection experiments lose sensitivity.

This leads to \textit{different} astrophysical targets being required to push to lower masses, such as planets. Due to their lower core temperatures, less kinetic energy can be transferred to DM during collisions, reducing the probability of escape. However, a planet's capacity to capture DM is limited, having a weaker gravitational potential, which implies they are less efficient at capturing DM than a star. Balancing these two leads to a narrow mass range, where planets, specifically Jupiter, can retain more light DM than the Sun~\cite{Leane:2021tjj, French:2022ccb, Blanco:2023qgi, Ansarifard:2024fan, Blanco:2024lqw}. We find that for Jupiter this range extends from $\sim1.2 - 1.3\GeV$, in very good agreement with ref.~\cite{Garani:2021feo}, up to $\sim3-4\GeV$. 

Here, we propose and calculate the first sensitivities of annihilating DM to neutrinos within Jupiter using existing and future neutrino observatories. We show that this analysis can probe \textit{lighter} masses than current solar bounds and even \textit{exceed} current bounds set by direct detection experiments.

Figure~\ref{fig:sketch} shows a sketch of the predicted sensitivities and compares them to current and future constraints set using the Sun. \textit{Our primary goal in this
Letter is to demonstrate searching for Jupiter-bound DM signals in neutrino experiments extends their physics reach and offers complimentary results to direct-detection experiments.}

This search adds to the rich field of using Jupiter as a dark matter probe, such as anomalous heating~\cite{Kawasaki:1991eu,Mack:2007xj,Adler:2008ky}, electron-trapping~\cite{Li:2022wix}, ionization~\cite{Blanco:2023qgi}, UV-airglow~\cite{Blanco:2024lqw}, and time-modulation~\cite{Ansarifard:2024fan}.

In the next section, we review the physics of neutrino fluxes from captured dark matter. There we show \textit{improved} calculations for DM capture in Jupiter. 
After that we present our sensitivity predictions for Super-Kamiokande (Super-K)~\cite{Super-Kamiokande:2002weg}, Hyper-Kamiokande (Hyper-K)~\cite{Hyper-Kamiokande:2018ofw, Hyper-Kamiokande:2022smq}, ORCA~\cite{KM3Net:2016zxf}, and the IceCube-Upgrade~\cite{Ishihara:2019aao}. Afterward we conclude. In the Supplemental Material, we provide further information.

\textbf{\textit{Neutrino Flux from DM annihilation.---}}
Jupiter, the largest and oldest planet in the solar system, is a gaseous planet. It is thought to be composed of a two-layer structure envelope enclosing a small rocky core. The envelope is mainly made of hydrogen and helium and also contains small traces of heavier elements. The outer envelope has a lower helium abundance than the inner envelope. 
Since we consider only DM-nucleon spin dependent interactions,   the hydrogen component in the planet's envelope is the sole target for DM scattering, and elements with no nuclear spin such as helium and heavier elements in the planet's core do not contribute to any scattering process. 

We assume that the DM accumulated in Jupiter annihilates solely to neutrinos with a yield per DM annihilation $dN_\nu/dE_\nu$, which in this case is just a delta function at the DM mass energy~\cite{Hooper:2002gs, Lindner:2010rr, Farzan:2011ck}. The neutrino flux also depends on the DM annihilation rate  $\Gamma_A$. This leads to a flux per neutrino flavor at a detector on Earth of
\begin{equation}
\dfrac{d\Phi_\nu}{dE_\nu} = \frac{\Gamma_A}{4\pi D_J^2 }    \dfrac{dN_\nu}{dE_\nu},
\end{equation}
where $D_J$ is the Earth-Jupiter distance. 

The annihilation rate, $\Gamma_A=AN_\chi^2/2$, is determined mainly by the number of particles,  $N_\chi$, accumulated in Jupiter at the present time and the annihilation cross section through 
\begin{equation}
A = \sigmav \frac{\int n_\chi^2(r) 4\pi r^2 dr}{\left(\int n_\chi(r) 4\pi r^2 \right)^2 dr},    
\end{equation}
where $n_\chi$ is the DM number density within Jupiter (see Supplemental Material). 
Here, we assume a thermal relic, i.e. $\sigmav=3\times10^{-26}\cm^3/\s$ and s-wave annihilation, as done when computing bounds from DM capture in the Sun at neutrino detectors~\cite{Super-Kamiokande:2004pou,Tanaka:2011uf, Super-Kamiokande:2015xms, Adrian-Martinez:2016gti, Aartsen:2016zhm, IceCube:2016dgk, ANTARES:2016xuh, IceCube:2021xzo, Gupta:2022lws}. 

The number of accreted DM particles at a given time, in turn, depends on three competing effects. (i) The rate at which they are captured by Jupiter's gravitational potential, $C$. (ii) Captured DM can scatter again with hydrogen atoms in Jupiter, gain energy and escape the planet before annihilating. This is called evaporation. (iii) The annihilation rate. Both, evaporation and annihilation deplete the number of accumulated DM particles. Then, the number of DM particles is obtained by solving the following differential equation:
\begin{equation}
\dfrac{dN_\chi}{dt} = C - E N_\chi - A N_\chi^2 ,
\label{eq:NWIMPs}
\end{equation}
where $E$ is the evaporation rate.  This equation has an exact analytical solution (see Supplemental Material), provided that capture, annihilation and evaporation rates remain constant throughout most of Jupiter's lifetime. At present time,  the number of DM particles in Jupiter's core is well approximated by
\begin{equation}
N_\chi \simeq \sqrt{\frac{C}{A}} \, \frac{1}{ \dfrac12 E \, t_\text{eq} +\sqrt{1+\left(\dfrac{E \, t_\text{eq}}{2}\right)^2}},    
\label{eq:Napprox}
\end{equation}
where $t_\text{eq} = 1/\sqrt{CA}$ is known as the capture-annihilation timescale. Note that when evaporation is negligible $N_\chi\simeq\sqrt{C/A}$ and $\Gamma_A\simeq C/2$. 

The spin-dependent capture rate is proportional to the scattering cross section $\sigma_{p\chi}^\text{SD}$, which we take to be a constant. 
To compute capture, evaporation, and annihilation rates in a consistent manner, we use a three-layer model of Jupiter that satisfies observational constraints~\cite{Nettelmann:2012}. 

\begin{figure}
    \centering
    \includegraphics[width=\columnwidth]{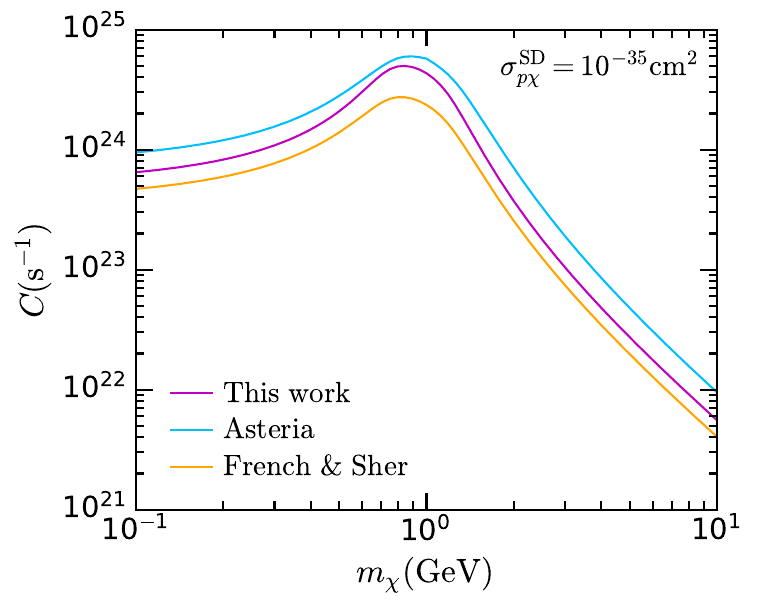}
    \caption{Spin dependent DM capture rate (magenta) for a DM-proton elastic scattering cross section $\sigma_{p\chi}^\text{SD}=10^{-35}\cm^2$. We also show results obtained with other estimations in the literature~\cite{French:2022ccb,Leane:2023woh} for comparison.}
    \label{fig:capture}
\end{figure}

In Fig.~\ref{fig:capture}, we show our computation of the capture rate for a DM-proton elastic scattering cross section $\sigma_{p\chi}^\text{SD}=10^{-35}\cm^2$ and compare our result in magenta with previous calculations in the literature \cite{French:2022ccb, Leane:2023woh}. The only difference between our calculation and others is that we have made use of a Jupiter model to properly account for the hydrogen number density in Jupiter's envelope as well as the escape velocity radial profile. The capture rate peaks around the proton mass due to resonance enhancement with hydrogen, as observed for various elements in the Earth~\cite{Gould:1987}. 

To calculate evaporation and annihilation rates we have followed refs.~\cite{Garani:2017jcj, Busoni:2017mhe}. We used the temperature profile provided by the aforementioned Jupiter model to estimate the DM density and velocity distribution in Jupiter's core once it has settled down in the center of the planet. 
Evaporation is relevant for light DM. In the particular case of Jupiter, this means DM of mass below $\sim 1.2 - 1.3 \GeV$ (in very good agreement with ref.~\cite{Garani:2021feo}), relevant for the range of scattering cross sections analyzed here (see Supplemental Material). Below this mass threshold captured DM is expected to evaporate before having the chance to annihilate into neutrinos. The exact value of the evaporation mass, as this threshold is known, can slightly vary around the above-mentioned values for a different assumption of annihilation cross section. 


\textbf{\textit{Neutrino Experiment Sensitivities.---}}
To estimate the feasibility of this analysis, we perform a rough estimate of the potential sensitivity: At $\sigma_{p\chi}^\text{SD} =  10^{-34}\;\mathrm{cm^2}$ and $m_\chi = 1\;\mathrm{GeV}$ the Jupiter capture rate (and annihilation rate) is $\sim 5\times 10^{25}\;\mathrm{s^{-1}}$. Assuming isotropic emission and scaling by the Jupiter-Earth sphere ($4\pi \times (8\times10^{13}\;\mathrm{cm})^2$), the resulting flux is then $\sim 5\times 10^{-4}\;\mathrm{(s\; cm^2)^{-1}}$. At 1 GeV (neutrino energy), the primary background is the atmospheric muon-neutrino flux $\sim 0.1\;\mathrm{(GeV\; s\; cm^2)^{-1}}$~\cite{Zhou:2023mou}. Using an angular resolution of 20$^\circ$ (the sky has $\sim 40000^{\circ^2}$) leads to a suppression of the atmospheric flux by $\sim 10^{-2}$. This puts the arriving neutrino flux from Jupiter ($5\times 10^{-4}\;\mathrm{(s\; cm^2)^{-1}}$), on par with the expected atmospheric flux, $ 10^{-3} \; \mathrm{(s\; cm^2)^{-1}}$. Since we are looking for a neutrino line, this estimate shows that we will search for an excess of events in Jupiter's direction centered at the DM mass.

Currently, the only detector with a significant set of GeV-scale neutrino events is Super-K~\cite{Super-Kamiokande:2002weg}. In the future JUNO~\cite{JUNO:2021vlw}, DUNE~\cite{DUNE:2015lol, DUNE:2020ypp}, and Hyper-K~\cite{Hyper-Kamiokande:2018ofw, Hyper-Kamiokande:2022smq} will have access to similar energy ranges. Here, we focus on Super-K and its successor, Hyper-K. Both are cylindrical water-based Cherenkov detectors with fiducial volumes of 22.5  and 187 kton, respectively. With angular resolution $\geq 20^\circ$~\cite{Super-Kamiokande:2005mbp}, detection efficiency $> 80\%$~\cite{Super-Kamiokande:2015qek} for CC $\nu_\mu$ events, and excellent energy resolution $\leq 10\%$~\cite{Shiozawa:1999sd, Drakopoulou:2017apf, Super-Kamiokande:2019gzr}, these detectors are ideal low-energy point source detectors.

Due to the dark matter evaporation mass $\sim 1\;\mathrm{GeV}$, we expect no neutrino signal events below 100 MeV. For this reason we restrict our analysis to neutrino energies $E_\nu \in [100\;\mathrm{MeV},\;5\;\mathrm{GeV}]$. For these energies, only atmospheric neutrinos will be a significant background. For the atmospheric neutrino background flux above 100 MeV, we use the HKKM11 model~\cite{Honda:2011nf}, which agrees well with Super-K measurements~\cite{Super-Kamiokande:2015qek}.

To predict the number of atmospheric and DM signal events in an energy bin $j$, we use
\begin{equation}
    \mathcal{N}^j = N_{t} \, \Delta t \int_{0}^{\infty} \mathrm{d}E_\nu \, 
    \frac{\mathrm{d}\Phi}{\mathrm{d}E_\nu}(E_\nu) \, \sigma_{j}(E_\nu)\epsilon (E_\nu),
\label{eq:yield}
\end{equation}
where $N_t$ is the number of target atoms (water), $\epsilon$ is the detection efficiency (80\% for Super-K and Hyper-K), and $\Delta t$ is the detector livetime.  For the exposure, we assume 15.9 (10) live time years for Super-K (and Hyper-K). $\sigma$ is the neutrino-oxygen cross section, which we get from {\tt GENIE 3.2.0} with tune G18\_10a\_02\_11b, which is based on a local Fermi-gas model and an empirical meson-exchange model~\cite{Andreopoulos:2009rq, Andreopoulos:2015wxa, GENIE:2021zuu}. Here we are using the neutrino-oxygen cross section as an approximation for the total neutrino-water cross section.

With 350 (1870) kton-years (15.9  and 10 years) of Super-K (Hyper-K) data, we predict $\mathcal{O}(25k)$ ($\mathcal{O}(100k)$) atmospheric muon neutrino events. 
These results agree well when scaled to SK data for 1489 days (SK-I), resulting in $\mathcal{O}(6k)$ events~\cite{Zhou:2023mou, Super-Kamiokande:2005mbp}.

To reduce this background, we apply angular cuts, defining a region of 20$^\circ$ around the direction of Jupiter and removing all events outside of it. With an angular reconstruction uncertainty of $\sim 20^\circ$~\cite{Super-Kamiokande:2019gzr}, this removes nearly 99\% of the atmospheric neutrino events, leaving only $\mathcal{O}(250)$ ($\mathcal{O}(1k)$) between 100 MeV and 5 GeV for Super-K (Hyper-K).

\begin{figure}
    \centering
    \includegraphics[width=0.5\textwidth]{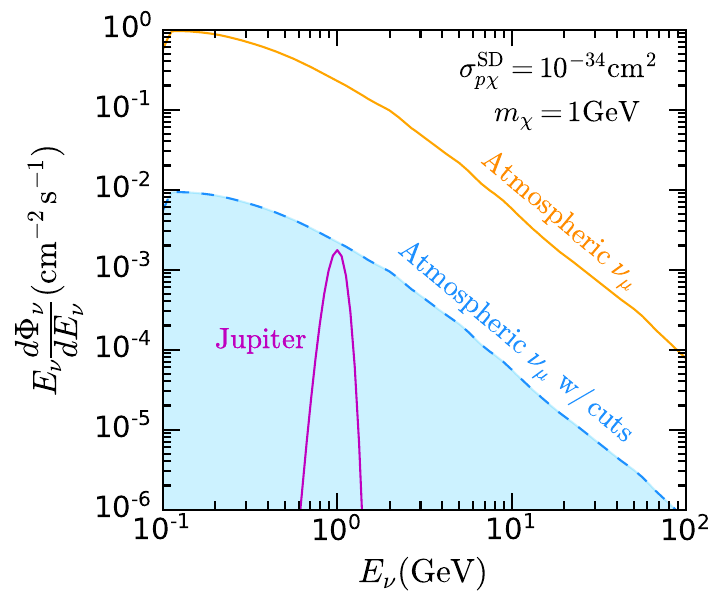} 
    \caption{Comparison of the atmospheric muon neutrino flux at Kamioka~\cite{Honda:2011nf} (orange) and the expected flux from DM annihilation (magenta). The dashed blue line shows the expected flux after directional cuts.}    
    \label{fig:flux_comp}
\end{figure}

In Fig.~\ref{fig:flux_comp}, we compare the arriving muon neutrino flux from Jupiter due to DM annihilation and the atmospheric background flux~\cite{Honda:2011nf} before and after directional cuts. The angular cuts cause a reduction of $\sim 100$. This causes the expected flux from Jupiter at $\sigma_{p\chi}^\text{SD} = 10^{-34}\;\mathrm{cm^2}$ to \textit{exceed} the expected background. For the fluxes shown, we have included a 10\% energy smearing.

Since the signal is a neutrino line, we can then apply an additional energy cut around the dark matter mass. We set this to be 10\% of the mass. This removes an additional 90\% of the atmospheric background. This means that for a given dark matter mass, we expect $\mathcal{O}(25)$ ($\mathcal{O}(100)$) background events. Thus, for a $2\sigma$ sensitivity, we need $\sim 10$ ($\sim 20$) signal events. This is equivalent to a cross section of $\sigma_{p\chi}^\text{SD} = 4\times 10^{-35}\;\mathrm{cm^2}$ ($\sigma_{p\chi}^\text{SD} = 1.5\times 10^{-35}\;\mathrm{cm^2}$) for Super-K (Hyper-K) at $m_\chi = $ 1 GeV.

Note that the final energy cut assumes a neutrino line. In other scenarios, such as the annihilation to $u\bar{u},\;\mu^+ \mu^-$ and $\tau^+ \tau^-$, such a sharp spectral feature at the dark matter mass would not exist~\cite{Liu:2020ckq}. Conservatively, assuming the energy cut is not possible in such scenarios, this would increase the atmospheric background by a factor of $\sim 10$, reducing the resulting sensitivity by $\sim\times 3$.

To calculate a more precise sensitivity, we generate 100k trials while accounting for a $\sim 20\%$ uncertainty on the neutrino-oxygen cross section. For each trial, we then construct a background model based on all neutrinos not coming from Jupiter (since we do not expect any DM signal). This effectively removes the cross section uncertainty. We then invert the standard frequentist hypothesis test~\cite{Feldman:1997qc} and set lower bounds for different dark matter masses on $\sigma^\mathrm{SD}_{p\chi}$.

\begin{figure}[t]
    \centering
    \includegraphics[width=\columnwidth]{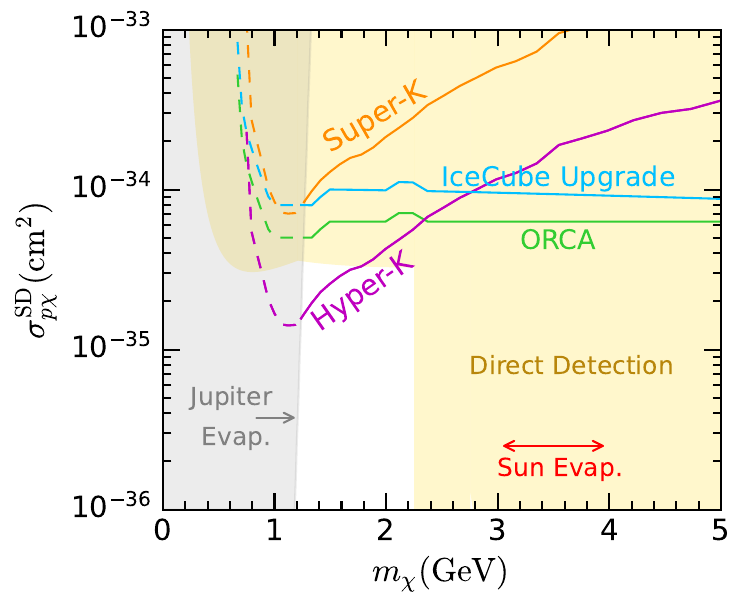}
    \caption{The 95\% sensitivities calculated here for Super-K, Hyper-K, ORCA and the IceCube Upgrade. We show the current direct detection constraints by NEWS-G~\cite{NEWS-G:2024jms}, CRESST-III~\cite{CRESST:2022dtl}, PICASSO~\cite{Behnke:2016lsk} and PICO-60~\cite{Amole:2019fdf} shaded in yellow. The gray region and the region within the right arrow
     show where evaporation becomes relevant in Jupiter and in the Sun, respectively.}
    \label{fig:bounds}
\end{figure}

Figure~\ref{fig:bounds} shows the resulting 95\% sensitivities for both Super-K and Hyper-K and compares them to current direct-detection constraints~\cite{NEWS-G:2024jms,CRESST:2022dtl,Behnke:2016lsk,Amole:2019fdf}. At $\sim 1$ GeV, we predict that Hyper-K, with 10 years of collected data, can exceed current direct-detection experiments, 
due to its larger volume. It is worth noting that NEWS-G limits outperform CRESST-III for $m_\chi \lesssim 1.2\;\mathrm{GeV}$, precisely the region where evaporation comes to play.  Note however that the upgraded version of the COSINE-100 experiment is expected to  outperform our Hyper-K projections, as very recently reported~\cite{COSINE-100:2025kbw}. 

We also show a sensitivity prediction for ORCA~\cite{KM3Net:2016zxf} using the angular resolution from ref.~\cite{KM3NeT:2020zod} and the effective area from ref.~\cite{Benfenati:2023uuz}. While ORCA's size means its yield for dark matter masses $\geq 1.5\;\mathrm{GeV}$ is higher, the larger angular reconstruction error ($\geq 30^\circ$)~\cite{Galata:2015gxa}, results in a far larger atmospheric background and a reduction in sensitivity. Additionally, ORCA (or other neutrino telescopes), needs to contend with an additional atmospheric background, miss-reconstructed atmospheric muons. Examples of rejection methods used to reduce this background are: boosted decision trees~\cite{Fusco:2015nky, KM3NeT:2025tiq} using outer layers of the detector as a veto~\cite{IceCube:2012jwm, IceCube:2013low, IceCube:2024fxo} or focusing on starting tracks within the detector~\cite{IceCube:2013low, IceCube:2024fxo}. Together, these methods have reduced contamination to below 10\%. For this reason, we assume the atmospheric muon background can be reduced to a negligible level for our sensitivity estimate. 

Of note is that recent machine-learning techniques show promise~\cite{KM3NeT:2020zod, Guderian:2022cco, Abbasi:2022ypr, Sogaard:2022qgg, PenaMartinez:2023rya} in improving ORCA's (and other neutrino telescope's) angular reconstruction, which would allow ORCA to exceed Hyper-K's sensitivity. 

We also give our prediction for the IceCube-Upgrade's~\cite{Ishihara:2019aao} sensitivity to muon neutrinos from dark matter annihilation coming from Jupiter. Using the effective area from ref.~\cite{Baur:2019jwm}, and assuming a similar angular reconstruction to ORCA's, the Upgrade's sensitivity is overall better than Super-K's and $\sim 60\%$ of ORCA's. 

Here, we exclusively used muon-neutrino events due to the ease of directional reconstruction of muon tracks. These results can easily be improved by including electron neutrino events, doubling the expected signal events.

\textbf{\textit{Conclusion and Discussion.---}}
Currently, the most stringent indirect bounds on GeV scale DM scattering off protons are set by observing the Sun. The reach of this method is naturally limited by the DM evaporation mass of the Sun.

Here, we proposed a new method to push these bounds to even lower DM masses by searching for neutrinos in the direction of Jupiter. These neutrinos are produced by captured DM annihilating within Jupiter. While the expected signal flux lies far below the atmospheric neutrino flux, Super-K's and Hyper-K's excellent angular reconstruction can suppress the atmospheric neutrino flux to such an extent that this new signal will appear as an \textit{excess} in neutrino events in the direction of Jupiter.

Using exclusively muon-neutrino events, we showed that Super-K can \textit{exceed} current direct detection constraints for dark matter interactions at approximately 1 GeV, while Hyper-K can improve on them even further. In the future, these sensitivities can be further improved by including electron-neutrino events. While the directional reconstruction is slightly worse~\cite{Super-Kamiokande:2019gzr}, the expected background is reduced~\cite{Honda:2011nf}, making this channel potentially better than the one chosen here.

\begin{acknowledgments}
\textbf{\textit{Acknowledgements.---}}
We would like to thank Carlos Blanco, Rebecca K. Leane, Marianne Moore, and Edward Kearns for their helpful and insightful comments.
SR  was partially supported by the UK STFC grant ST/T000759/1 and by the Fermi National Accelerator Laboratory (Fermilab), a U.S.
Department of Energy, Office of Science, HEP User Facility. 
SR thanks the  Institute for Nuclear Theory at the University of Washington for its hospitality and the Department of Energy for partial support during the completion of this work.
This work was performed in part at Aspen Center for Physics, which is supported by National Science Foundation grant PHY-2210452. 
SR  acknowledges CERN TH Department for its hospitality while this research was being carried out. 
SMB was supported by the Australian Research Council through Discovery Project DP220101727 plus the University of Melbourne’s Research Computing Services and the Petascale Campus Initiative.
\end{acknowledgments}
\bibliography{references}
\clearpage
\onecolumngrid
\appendix*
\setcounter{equation}{0}
\section{\Large Supplemental Material}
\label{sec:suppl}
\vspace*{0.2cm}

Here, we provide additional details that may be helpful. We discuss capture rate calculations, evaporation rates, annihilation rates, and flux comparisons.
\subsection{Capture rate}

The  DM  capture rate in the Earth was first calculated by Gould~\cite{Gould:1987}. A similar approach can be adopted for Jupiter, i.e. 
\begin{equation}
    C = 4\pi\frac{\rho_\chi}{m_\chi}\int_0^{\RJup} dr \,  r^2 
    \int_0^{\infty} du_\chi \frac{w(r)}{u_\chi} \fMB(u_\chi) \Omega^-(w),\label{eq:capdef}
\end{equation}
where $\fMB(u_\chi)$ is the DM velocity distribution far from the Solar system, that we have taken to be Maxwell-Boltzmann (MB) and defined as in ref.~\cite{Busoni:2017mhe}, $w(r)=\sqrt{u_\chi^2+\vesc^2(r)}$ is the DM speed at a radial distance $r$ from the planet before the collision takes place, and $\vesc$ is the escape velocity.
To calculate the scattering rate  $\Omega^-$, we consider spin-dependent interactions parameterized by a constant DM-proton cross section $\sigma_{p\chi}^\text{SD}$. This is the only available targets are hydrogen atoms. 
In addition, in the DM mass regime we are interested in, we can safely neglect the thermal motion of the targets, whose temperature can reach at most  ${\cal O}(1\text{eV})$ in the center of the planet \cite{Nettelmann:2012}. Thus, the DM scattering rate from  a velocity $w(r)$ to a velocity lower than the escape velocity is given by \cite{Busoni:2017mhe}
\begin{equation}
    \Omega^-(w) = \frac{\mu_+^2}{\mu w}\sigma_{p\chi}^\text{SD}  \nH(r) \left( \vesc^2(r)- w^2\frac{\mu_-^2}{\mu_+^2}  \right)  \Theta\left( \vesc^2(r)- w^2\frac{\mu_-^2}{\mu_+^2}  \right), \label{eq:scattrate}
\end{equation}
where $\mu=m_\chi/m_\text{H}$, $\mu_\pm=(\mu\pm1)/2$ and $n_H(r)$ is the hydrogen number density.

From Eqs.~\ref{eq:capdef} and \ref{eq:scattrate}, the capture rate scales linearly with  $\sigma_{p\chi}^\text{SD}$. However, if the cross section is larger than a certain threshold value, the capture rate will begin to saturate to its maximum value, the so-called geometric limit. In this limit all DM particles traversing the planet are captured regardless of the value of $\sigma_{p\chi}^\text{SD}$. To account for this, we introduce an optical factor $\eta(r)$ in Eq.~\ref{eq:capdef}, whose purpose is to remove captured DM particles from the incoming DM flux, and is in principle a function of the optical depth $\tau_\chi$ seen by a DM particle as it goes through the planet's envelope~\cite{Gould:1992}
\begin{equation}
    \eta(\tau_\chi)=e^{-\tau_\chi}.
\end{equation}
To calculate the optical  depth at a specific radial position we use the approach of refs.~\cite{Gould:1992,Busoni:2017mhe,Hong:2024ozz}. 
It is noteworthy that ref.~\cite{French:2022ccb} assumed the geometric cross section to be $10^{-34}\cm^2$ for the same DM mass range we have considered, while we find it varies in the range $\sim 3\times 10^{-33} - 10^{-31}\cm^2$. 
\subsection{Evaporation rate}
As the capture process, the evaporation rate has also been previously studied, initially in the context of the Sun~\cite{Gould:1987b,Gould:1990,Garani:2017jcj,Busoni:2017mhe},  which is the relevant regime for Jupiter
\begin{equation}
E = \int_0^{\RJup} dr\, 4\pi r^2 \eta(r) n_\chi(r) \int_0^{\vesc(r)} dw \,  4\pi w^2 f_\chi(w,r) \ \Omega^+(w),    
\end{equation}
where $n_\chi$ and $f_\chi$ are the DM number density and velocity distribution within Jupiter, respectively.  We calculate the latter using an interpolation between the isothermal (iso) and local thermodynamical equilibrium (LTE) regimes~\cite{Bottino:2002pd,Scott:2008ns,Garani:2017jcj,Busoni:2017mhe,Banks:2021sba}, i.e. 
\begin{equation}
 n_\chi(r) f_\chi(w,r) = [1-f(K)] \, n_\chi^{\rm iso}(r)  \left(\frac{m_\chi}{2\pi T_\chi}\right)^{3/2} \exp\left[-\frac{m_{\chi}w^2}{2T_\chi}\right]  + f(K) \, n_\chi^{\rm LTE}(r)  \left(\frac{m_\chi}{2\pi \TJup(r)}\right)^{3/2} \exp\left[-\frac{m_{\chi}w^2}{2\TJup(r)}\right].
\end{equation}
where $T_\chi$ is the temperature of the DM isothermal distribution as defined in ref.~\cite{Busoni:2017mhe}, $T_J(r)$ is Jupiter's temperature at a given radial shell $r$, and
\begin{equation}
    f(K) \approx \frac{1}{1+(K/K_0)^{1/\tau}}, 
\end{equation}
where $K_0=0.4$ and $\tau=0.5$ for the Sun~\cite{Bottino:2002pd,Scott:2008ns}. $K$ is the Knudsen number calculated as the ratio of the mean free path $\ell=1/(n_H\sigma_{p\chi})$ to the radius of the DM virialized distribution $r_\chi$
\begin{equation}
K=\frac{\ell}{r_\chi}, \qquad r_{\chi}=\sqrt{\frac{3 \TJup(0)}{2\pi G\rho_c m_\chi}}, 
\end{equation}
with $\rho_c$ the density at  Jupiter's center. The expressions for $n_\chi^{\rm iso}$ (large mean free path) and $n_\chi^{\rm LTE}$ (short mean free path) can be found in refs.~\cite{Garani:2017jcj,Busoni:2017mhe}. 

Finally, the up-scattering rate $\Omega^+$, which is the probability of captured DM to gain energy in subsequent collisions and achieve a velocity greater than the escape velocity, reads
\begin{equation}
 \Omega^{+}(w) = \frac{16 \mu_{+}^4}{\sqrt{\pi}} \sigma_{p\chi}^\text{SD}\int_{\vesc(r)}^\infty  d v  \int_0^\infty ds \int_0^\infty d t' \, \left[\frac{m_p}{2T_J(r)}\right]^{3/2} \nH(r) \frac{v t'}{w} \exp\left[-\frac{m_p}{2T_J(r)}v_T^2\right] \Theta(t'+s-v)\Theta(w-|t'-s|), 
\end{equation}
where $v_T^2 = 2\mu\mu_+ (t')^2+2\mu_+ s^2-\mu w^2$ is the target squared velocity before the collision in the center of mass (c.m.), $s$ is the c.m. velocity, and $t'$ the DM initial speed also in the c.m. frame. 

\subsection{Annihilation rate}
\begin{figure}
    \centering
    \includegraphics[width=0.49\textwidth]{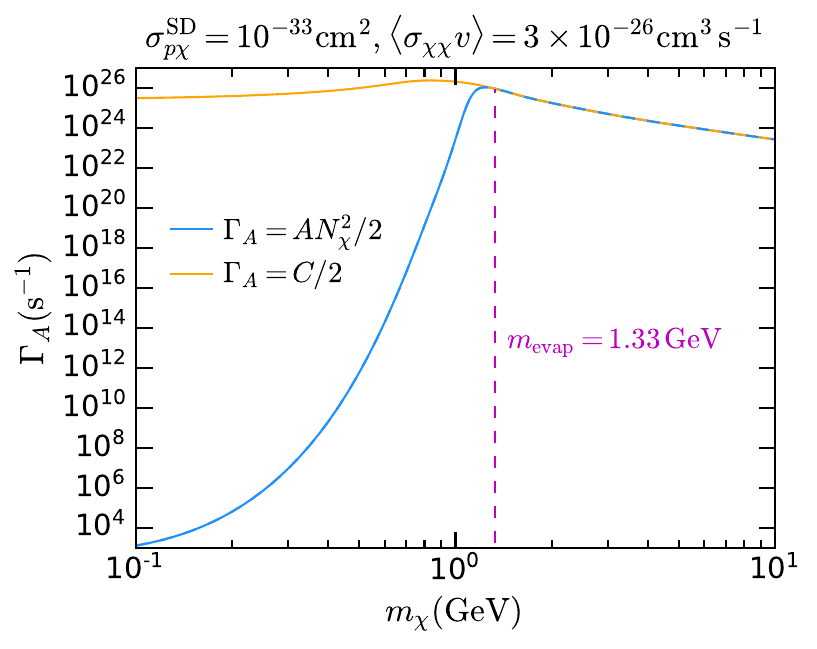} 
    \includegraphics[width=0.49\textwidth]{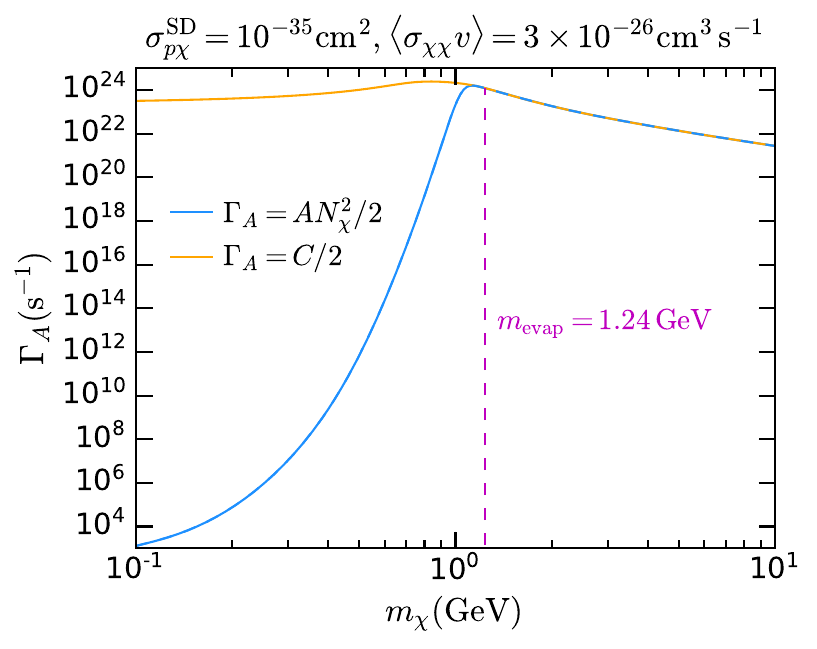}
    \caption{Illustration of the calculation of the evaporation mass. The exact computation of the annihilation rate is shown in blue, and the corresponding approximation when evaporation is negligible is depicted in orange. The evaporation mass for different values of $\sigma_{p\chi}^\text{SD}$ is shown in magenta. 
    }    
    \label{fig:mevap}
\end{figure}

To calculate the annihilation rate, specifically the factor $A$, we have  assumed s-wave annihilation \cite{Busoni:2013kaa,Garani:2017jcj}, which leads to
\begin{equation}
A = \sigmav \frac{\int n_\chi^2(r) 4\pi r^2 dr}{\left(\int n_\chi(r) 4\pi r^2 \right)^2},    
\end{equation}
with the DM number density in Jupiter calculated again using the fit $f(K)$
\begin{equation}
 n_\chi(r)  = [1-f(K)] \, n_\chi^{\rm iso}(r)    + f(K) \, n_\chi^{\rm LTE}(r).  
\end{equation}

For completeness, we give here the exact solution to the equation for the number of DM particles in Jupiter, Eq.~\ref{eq:NWIMPs}
\begin{equation}
N_\chi(t) = \sqrt{\frac{C}{A}} \, \left[\frac{\tanh(\beta \, t/t_\text{eq})}{\beta + \frac12 E \, t_\text{eq} \tanh(\beta \, t/t_\text{eq})}\right],    
\label{eq:Nchi}
\end{equation}
where
\begin{equation}
\beta = \sqrt{1+\left(\frac{E \, t_\text{eq}}{2}\right)^2}. 
\end{equation}
Note than when $\beta \, t \gg t_\text{eq}$, we obtain the approximation given in Eq.~\ref{eq:Napprox}.

Finally, once we have calculated capture, evaporation and annihilation rates, we can estimate the evaporation mass $m_\text{evap}$. This is done by requiring the approximation of the annihilation when evaporation is negligible $\Gamma_A=C/2$ to be $99\%$ of the exact expression  $\Gamma_A=A N_\chi^2/2$, with $N_\chi$ given by Eq.~\ref{eq:Nchi} and $t=3.4 \, \text{Gyr}$, a Gyr less than the actual Jupiter's age~\cite{Nettelmann:2012} to account for any possible change in the planet's structure, composition and temperature that could lead to different values of the capture, evaporation and annihilation rates. 
We showcase this estimation in Fig.~\ref{fig:mevap}. It is evident that we cannot use the approximation $\Gamma_A=C/2$, for DM masses below $m_\text{evap}$, i.e. the sub-GeV regime, since evaporation makes the annihilation rate dramatically be suppressed by several orders of magnitude. 
It is worth noting than due to the values of the cross sections assumed in Fig.~\ref{fig:mevap}, the evaporation rate was calculated in the optically thin limit, i.e. with $\eta(r)\rightarrow1$, as well as the evaporation mass curves for Jupiter and the Sun in Fig.~\ref{fig:sketch}. The optical factor $\eta(r)$ causes the evaporation mass to diminish at large cross sections which is not relevant for the sensitivities  shown in this work. 

\end{document}